\documentclass[twocolumn,showpacs,preprintnumbers,amsmath,amssymb]{revtex4}

\usepackage{graphicx}
\usepackage{dcolumn}
\usepackage{bm}
\usepackage{amsthm}

\def\PP{\mathbb{P}}
\def\tr{{\rm tr}\, }
\newcommand{\begth}{\begin{theorem}}
\newcommand{\enth}{\end{theorem}}
\newcommand{\bpr}{\begin{proof}}
\newcommand{\epr}{\end{proof}}
\newcommand{\be}{\begin{equation}}
\newcommand{\ee}{\end{equation}}
\newcommand{\bes}{\begin{equation*}}
\newcommand{\ees}{\end{equation*}}
\newcommand{\bea}{\begin{eqnarray}}
\newcommand{\eea}{\end{eqnarray}}
\newcommand{\beas}{\begin{eqnarray*}}
\newcommand{\eeas}{\end{eqnarray*}}
\newcommand{\non}{\nonumber}

\newtheorem{theorem}{Theorem}
\begin{document}


\title{The classical capacity of quantum channels with memory\\}

\author{Tony Dorlas}
\email{dorlas@stp.dias.ie}
\author{Ciara Morgan}%
 \email{cmorgan@stp.dias.ie}
\affiliation{%
School of Theoretical Physics,
Dublin Institute for Advanced Studies, 10 Burlington Road, Dublin
4, Ireland.
}%


\begin{abstract}
We investigate the classical capacity of two quantum channels with
memory: a periodic channel with depolarizing channel branches, and
a convex combination of depolarizing channels. We prove that the
capacity is additive in both cases. As a result, the channel
capacity is achieved without the use of entangled input states. In
the case of a convex combination of depolarizing channels the
proof provided can be extended to other quantum channels whose
classical capacity has been proved to be additive in the
memoryless case.
\end{abstract}

\maketitle

\section{Introduction\protect\\}

The problem of determining the classical information-carrying
capacity of a quantum channel is one which has not been fully
resolved to date. In the case where the input to the channel is
prepared in the form of non-entangled states, the classical
capacity can be determined using a simple formula. However, if
entanglement between multiple uses of the channel is permitted,
then the channel capacity can only be determined asymptotically.
We now discuss these concepts in detail.

Using product-state encoding, i.e. when a message is encoded into
a tensor product of $n$ quantum states on a finite-dimensional
Hilbert space $\cal H$, this state can then be transmitted over a
quantum channel given by a completely positive trace-preserving
map (CPT) $\Phi^{(n)}$ on ${\cal B}({\cal H}^{\otimes n})$. The
associated capacity is known as the product-state capacity of the
channel.
Note, that a channel is said to be memoryless if the noise acts
independently on each state sent over the channel, i.e.
if $\Phi^{(n)} = \Phi^{\otimes n}$ is a memoryless
channel, then the product-state capacity is given by the supremum of
the Holevo $\chi$-quantity given by the right-hand side of
(\ref{holevo-bound}), evaluated over all possible input state
ensembles. This is also known as the Holevo capacity
$\chi^*(\Phi)$ of the channel.

Memoryless quantum channels have received a great deal of attention.
However, such channels which have no correlation between noise acting on
successive channel inputs can be seen to be unrealistic, since real-world
quantum channels may not exhibit this independence and correlations between errors are common.
Noise correlations are also necessary for certain models of quantum communication
(see \cite{Bose03}, for example). These correspond to quantum memory channels.

On the other hand, a block of input states could be permitted to
be entangled over $n$ channel uses. The classical capacity is
defined as the limit of the capacity for such $n$-fold entangled
states divided by $n$, as $n$ tends to infinity. If the Holevo
capacity of a memoryless channel is additive, then it is equal to
the classical capacity of that channel and there is no advantage
to using entangled input state codewords. The additivity
conjecture for the Holevo capacity of most classes of memoryless
channel remains open. However, the classical capacity of certain
memoryless quantum channels have been shown to be additive: see
\cite{KingUnital}, \cite{Shor02}, \cite{King03}, for example. On
the other hand, there now exists an example of a memoryless
channel for which the conjecture does not hold: see \cite{Hastings}.

We remark that, Shor \cite{Shor04} (see also Fukuda \cite{Fukuda07})
proved that the additivity conjectures involving the entanglement
of formation \cite{BDiVSWK}, the minimum output entropy \cite{KR01}, the
strong superadditivity and the Holevo capacity \cite{Hol98},
\cite{SW97} are in fact equivalent.

In this article we consider the classical capacity of two
particular types of channels with \emph{memory} consisting of depolarizing
channel branches, namely a periodic channel and a convex
combination of memoryless channels.

In \cite{DD07Per} Datta and Dorlas derived a general expression
for the classical capacity of a quantum channel with
\emph{arbitrary} Markovian correlated noise. We consider two
special cases of this channel, that is, a periodic channel with
depolarizing channel branches and a convex combination of
memoryless channels, and we prove that the corresponding
capacities are additive in the sense that they are equal to the
product-state capacities. A convex combination of memoryless
channels was discussed in \cite{DD07CC} and can be described by a
Markov chain which is aperiodic but not irreducible. Both channels
are examples of a channel with long-term memory.

The article is organized as follows. The objectives as discussed
above are formalized in Section \ref{prelim}. In Section II we
introduce the periodic channel and investigate the product state
capacity of the channel with depolarizing channel branches. We
derive a result based on the invariance of the maximizing ensemble
of the depolarizing channel, which enables us to prove that the
capacity of such a periodic channel is additive. In Section III the additivity of the classical capacity
of a convex combination of depolarizing channels is proved. The is
done independently of the result derived in Section II and can
therefore be generalized to a class of other quantum channels.

\subsection{Preliminaries}\label{prelim}

A quantum state is described by a positive operator of unit trace
$\rho \in \mathcal{B}\left(\mathcal{H} \right)$, where
$\mathcal{B}\left(\mathcal{H} \right)$ denotes the algebra of
linear operators acting on a finite dimensional Hilbert space
$\mathcal{H}$. The transmission of classical information over a
quantum channel is achieved by encoding the information as quantum
states. To accomplish this, a set of possible input states $\rho_j
\in \mathcal{B}\left(\mathcal{H}\right)$ with probabilities $p_j$
are prepared, describing the ensemble $\{p_j,\rho_j\}$. The
average input state to the channel is expressed as $\rho = \sum_j
p_j \rho_j$. For a channel given by a completely positive trace
preserving map $\Phi:{\mathcal{B}}({\cal H}) \to
{\mathcal{B}}({\cal K})$, the average \emph{output} state is
$\tilde{\rho} = \sum_j p_j \Phi(\rho_j)$.

When a state is sent though a noisy quantum channel, the amount of
information about the input state that can be inferred from the
output state is called the \emph{accessible information}. The
Holevo bound, \cite{Hol73}, provides an upper bound on the
accessible information and is given by,
\be\label{holevo-bound}
H\left(X:Y\right) \leq S\left(\sum_j p_j \, \rho_{j}\right) -
\sum_j p_j \, S\left(\rho_{j}\right), \ee
where $S(\rho) = -\tr \left( \rho \, \log \,
\rho\right)$ is the von Neumann entropy. Note that all logarithms are taken to the base $2$. Here $X$ is the random variable representing the
classical input to the channel. The possible values $x_j$ are
mapped to states $\rho_j$ which are transformed to $\Phi(\rho_j)$
by the channel. Then, a generalized measurement with corresponding
Positive Operator-Valued Measure (POVM) $\{E_j\}$ allows the determination of the output random
variable $Y$ with conditional probability distribution given by
\be \PP(Y=x_k\,|\,X=x_j) = \tr (\Phi(\rho_j) E_k). \ee The right
hand side of equation (\ref{holevo-bound}) is called the
Holevo-$\chi$-quantity, usually denoted $\chi \left(
\{p_j,\Phi(\rho_j)\} \right)$. Holevo \cite{Hol98} and Schumacher
and Westmoreland \cite{SW97} proved independently that for a
memoryless channel, the upper bound on $H \left(X:Y \right)$ is
asymptotically achievable. Using product-state coding as described
above, the input message to the channel is encoded into a product
state codeword of length $n$ and is transmitted over $n$ copies of
the channel. The Holevo Schumacher Westmoreland (HSW) Theorem
states that the product state capacity of that channel is given by
the supremum, over all input ensembles, of the Holevo quantity of
that channel, where each input state is prepared as a product
state codeword. In other words, the rate at which classical
information can be sent over a quantum channel, where each input
codeword is a product state comprised of states belonging to an
ensemble is given by the following ``single-letter'' formula,
\be\label{HSW} \chi^*\left(\Phi\right) =
\sup_{\{p_j,\rho_{j}\}}\left[ S \left(\Phi\left( \sum_j p_j\rho_j
\right)\right) - \sum_j p_jS\left(\Phi(\rho_j)\right)\right] \ee
where $S$ is the von Neumann entropy.
An ensemble which maximizes the Holevo quantity $\chi$ of a
channel is known as a maximizing or optimal ensemble.

It was first shown in \cite{FBS96} that for some channels, it is
possible to gain a higher rate of transmission by sending
entangled states across multiple copies of a quantum channel. In
general, allowing both entangled \emph{input} states and output
measurements and with an unlimited number of copies of the
channel, the classical capacity of $\Phi$ is given by \cite{Hol72}
\be\label{ult} C \left( \Phi \right) = \lim_{n \rightarrow
\infty} \frac{1}{n} \, \chi^* \left( \Phi^{(n)}\right), \ee where
\beas \chi^*(\Phi^{(n)}) =
\sup_{\{p_j^{(n)},\rho_{j}^{(n)}\}} \biggl[ S \left(\Phi^{(n)}\left(
\sum_j p_j^{(n)}\rho_j^{(n)} \right)\right) \\ - \sum_j
p_j^{(n)}S\left(\Phi^{(n)}\left(\rho_j^{(n)}\right)\right)\biggr]
\eeas
denotes the  Holevo capacity of the channel $\Phi^{(n)}$ with
an $n$-fold input state ensemble.

The Holevo capacity of a channel $\Phi$ is said to be
\emph{additive} if the following holds for an arbitrary channel
$\Psi$: \be \chi^*\left( \Phi \otimes \Psi \right) = \chi^*\left(
\Phi \right) + \chi^*\left( \Psi \right). \ee

In particular, if we can prove that the Holevo capacity of a
particular channel is additive then \be \chi^* \left(
\Phi^{\otimes n} \right) = n \; \chi^*\left( \Phi \right), \ee
which implies that the classical capacity of that channel is equal
to the product state capacity, that is,  \bea\label{classCap} C
\left( \Phi \right) = \chi^*\left( \Phi \right). \eea This will
imply that the classical capacity of that channel cannot be
increased by entangling inputs across two or more uses of the
channel. Additivity has been proved for unital qubit channels
\cite{KingUnital}, entanglement-breaking channels \cite{Shor02},
and the depolarizing channel \cite{King03}. Here we use the latter
result to prove equation (\ref{classCap}) for a periodic channel
with depolarizing channel branches and for a convex combination of
depolarizing channels.

\section{The periodic channel\protect\\}

A periodic channel acting on an $n$-fold density operator has the
form \be\label{periodic} \Omega^{\left(n \right)} \left( \rho
^{\left(n \right)} \right) = \frac{1}{L} \sum_{i=0}^{L-1} \left(
\Omega_i \otimes \Omega_{i+1} \otimes \cdots\otimes \Omega_{i+n-1}
\right) \left( \rho ^{\left(n \right)} \right), \ee
where $\Omega_i$ are CPT maps and the index is cyclic modulo the period $L$.

We denote the Holevo quantity for the $i$-th branch of the channel
by $\chi_i(\{p_j,\rho_j\})$, i.e. \be\label{Hquantity}
\chi_i(\{p_j, \rho_j\}) = S \left( \sum_j p_j \Omega_i\left(
\rho_j \right)\right) - \sum_j p_j S\left(\Omega_i(\rho_j)\right).
\ee

Since there is a correlation between the noise affecting
successive input states to the periodic channel (\ref{periodic}),
the channel is considered to have memory and the product state
capacity of the channel is no longer given by the supremum of the
Holevo quantity. Instead, the product state capacity of this
channel is given by the following expression \be\label{cap_Per}
C_p \left( \Omega \right) = \frac{1}{L} \sup_{\{p_j, \rho_j\}}
\sum_{i=0}^{L-1} \chi_i(\{p_j, \rho_j\}). \ee Next, we introduce
the depolarizing channel and investigate the product state
capacity of a periodic channel with depolarizing channel branches.

\subsection{A periodic channel with depolarizing channel
branches}\label{sect_Dep}

The quantum depolarizing channel can be written as follows \be
\Delta_{\lambda} \left( \rho \right)= \lambda \rho + \frac{1 -
\lambda}{d} I \ee where $\rho \in \mathcal{B}\left( \mathcal{H}
\right)$ and $I$ is the $d \times d$ identity matrix. Note that in
order for the channel to be completely positive the parameter
$\lambda$ must lie within the range \be - \frac{1}{d^2 -1} \leq
\lambda \leq 1. \ee Output states from this channel have
eigenvalues $\left(\lambda + \frac{1-\lambda}{d}\right)$ with
multiplicity $1$ and $\left(\frac{1-\lambda}{d}\right)$ with
multiplicity $d-1$.

The minimum output entropy of a channel $\Phi$  is defined by \be
S_{min} \left( \Phi \right) = \inf_{\rho} S\left(\Phi \left( \rho
\right) \right). \ee It is easy to see that the product-state capacity
of the depolarizing channel is given by \be\label{depCap}
\chi^*\left( \Delta_{\lambda} \right) = log\left( d\right) -
S_{min} \left( \Delta_{\lambda} \right), \ee where the minimum
entropy is attained for any set of orthonormal vector states, and
is given by
\bea
 S_{min} \left( \Delta_{\lambda} \right) &=& - \left(
\lambda + \frac{1-\lambda}{d}\right) \, \log\left( \lambda +
\frac{1-\lambda}{d} \right) \non \\ &-& (d-1) \left(
\frac{1-\lambda}{d}\right) \, \log
\left(\frac{1-\lambda}{d}\right).
\eea

Next we show that the product state capacity of a periodic channel
with $L$ depolarizing channel branches is given by the sum of the
maximum of the Holevo quantities of the individual depolarizing
channels, in other words we show that
\be\label{conj} \frac{1}{L}
\sup_{\{p_j, \rho_j\}} \sum_{i=0}^{L-1} \chi_i(\{p_j,
\rho_j\})  = \frac{1}{L} \sum_{i=0}^{L-1}
\sup_{\{p_j, \rho_j\}} \chi_i(\{p_j, \rho_j\}). \ee
Let
$\Delta_{\lambda_1}, \Delta_{\lambda_2}, \cdots,
\Delta_{\lambda_L}$ denote $d$-dimensional depolarizing channels
with respective error parameters $\lambda_1, \lambda_2, \cdots,
\lambda_L$. Using the capacity given by equation (\ref{depCap})
and since every depolarizing channel can be maximized using a
single ensemble of orthogonal pure states independently of the
error parameter, the right-hand side of equation (\ref{conj}) can
be written as
\bea\label{sumChi} \frac{1}{L} \sum_{i=0}^{L-1}
\sup_{\{p_j, \rho_j\}} \chi_i(\{p_j, \rho_j\})  = 1 - \frac{1}{L}  \biggl[ S_{min}\left( \Delta_{\lambda_1}\right)  + \non \\ \cdots  +
S_{min}\left( \Delta_{\lambda_L} \right)\biggr]. \eea

Clearly, the left-hand side of equation (\ref{conj}) is bounded
above by the right-hand side. On the other hand, choosing the
ensemble to be an orthogonal basis of states with uniform
probabilities, we have \be \frac{1}{L} \sum_{i=0}^{L-1} \chi_i
(\{p_j, \rho_j\})  = 1 - \frac{1}{L} \sum_{i=0}^{L-1} S_{min}
\left( \Delta_{\lambda_i}\right). \ee We can now conclude that
equation (\ref{conj}) holds for a periodic channel with $L$
depolarizing branches of arbitrary dimension.

\subsection{The classical capacity of a periodic channel}

We now consider the classical capacity of the periodic channel,
$\Omega_{per}$ given by equation (\ref{periodic}), where $\Omega_i =
\Delta_{\lambda_i}$ are depolarizing channels with dimension $d$.
Denote by $\Psi_0^{(n)} , \dots, \Psi_{L-1}^{(n)}$ the following
product channels \bea\label{psi} \Psi_i^{(n)} = \Delta_{\lambda_i}
\otimes \dots \otimes \Delta_{\lambda_{i+n-1}}, \eea where the
index is taken modulo $L$.

We define a \emph{single use} of the periodic channel,
$\Omega_{per}$, to be the application of one of the depolarizing
maps $\Delta_{\lambda_i}$. If $n$ copies of the channel are
available, then with probability $\frac{1}{L}$ one of the product
branches $\Psi_i^{(n)}$ will be applied to an $n$-fold input
state.

We aim to prove the following theorem. \begth\label{theorem1} The
classical capacity of the periodic channel $\Omega_{per}$ with
depolarizing channel branches is equal to its product state
capacity, \beas C \left( \Omega_{per} \right) = \; C_p \left(
\Omega_{per} \right) = 1-\frac{1}{L} \sum_{i=0}^{L-1}
S_{min}(\Delta_{\lambda_i}). \eeas \enth

To prove Theorem \ref{theorem1} we first need a relationship
between the supremum of the Holevo quantity $\chi^*$ and the
channel branches $\Psi_i^{(n)}$. King \cite{King03} proved that
the supremum of the Holevo quantity of the product channel
$\Delta_{\lambda} \otimes \Psi$ is additive, where
$\Delta_{\lambda}$ is a depolarizing channel and $\Psi$ is a
completely arbitrary channel, i.e., \be\label{addpol} \chi^*
\left( \Delta_{\lambda} \otimes \Psi \right) = \chi^* \left(
\Delta_{\lambda} \right) + \chi^*\left( \Psi \right). \ee It
follows immediately that
\bea\label{add} \chi^* \left(
\Psi_i^{(n)} \right) &=& \chi^* \left( \Delta_{\lambda_i} \right) +
\chi^* \left( \Psi_{i+1}^{(n-1)} \right) \non \\ &=& \sum_{i=0}^{L-1} \chi^*
\left( \Delta_{\lambda_i} \right) + \chi^* \left( \Psi_i^{(n-L)}
\right). \eea
Next, we use this result to prove Theorem 1.

\begin{proof}
The classical capacity of an arbitrary quantum channel $\Omega$ is
given by \be\label{asympCap} C \left( \Omega \right) = \lim_{ n
\rightarrow \infty}\; \frac{1}{n} \, \sup_{\{ p_j^{(n)}, \, \rho_j^{(n)} \}} \chi \left( \{p_j, \Omega^{(n)} \left( \rho_j^{(n)}
\right) \} \right). \ee In Section \ref{sect_Dep} we showed that
the product state capacity of the periodic channel $\Omega_{per}$
can be written as \be C_p \left( \Omega_{per} \right) =
\frac{1}{L} \sum_{i=0}^{L-1} \chi^* \left( \Delta_{\lambda_i}
\right). \ee Using the product channels $\Psi_i^{(n)}\left(
\rho_j^{(n)}\right)$ defined by equations (\ref{psi}), the
periodic channel $\Omega_{per}$ can be written as
\bea\label{perChannel} \Omega_{per}^{(n)} \left( \rho_j^{(n)}
\right) = \frac{1}{L} \sum_{i=0}^{L-1} \Psi_i^{(n)}\left(
\rho_j^{(n)} \right). \eea

Since it is clear that \be\label{greatEq} C \left( \Omega_{per}
\right) \geq C_p \left( \Omega_{per}\right), \ee we concentrate
on proving the inequality in the other direction.

First suppose that \be\label{assumption} C \left( \Omega_{per}
\right) \geq \frac{1}{L} \sum_{i=0}^{L-1}  \chi^* \left(
\Delta_{\lambda_i} \right) + \epsilon, \ee
for some $\epsilon
>0$. Then $\exists n_0$ such that if $n \geq n_0$, then \be
\frac{1}{n} \, \sup_{\{  p_j^{(n)}, \, \rho_j^{(n)} \}} \chi \left(
\Omega_{per}^{(n)} \left( \rho_j^{(n)} \right)\right) \geq
\frac{1}{L} \sum_{i=0}^{L-1}  \chi^* \left( \Delta_{\lambda_i}
\right) + \frac{\epsilon}{2}. \ee The supremum in equation
(\ref{asympCap}) is taken over all possible input ensembles ${\{
\rho_j^{(n)}, \, p_j^{(n)} \}}$. Therefore, for $n \geq n_0$,
there exists an ensemble ${\{ \rho_j^{(n)}, \, p_j^{(n)} \}}$ such
that \be\label{ineqEn} \frac{1}{n} \,  \chi \left( \left\{ p_j,
\Omega_{per}^{(n)} \left( \rho_j^{(n)} \right) \right\} \right)
\geq \frac{1}{L} \sum_{i=0}^{L-1}  \chi^* \left(
\Delta_{\lambda_i} \right) + \frac{\epsilon}{2}. \ee  The Holevo
quantity can be expressed as the average of the relative entropy
of the average ensemble state with respect to members of the
ensemble \be \chi\left( \{ p_k,\, \rho_k \}\right) = \sum_k \,
p_k \, S \left( \rho_k , \,\big|\big|\, \sum_k \, p_k \,
\rho_k\right), \ee where, $S\left(A \,||\,B \right) = \tr \left(
A \, \log A \right) - \tr \left(A \, \log B\right)$, represents
the relative entropy of $A$ with respect to $B$. (Vedral
\cite{Vedral02} has argued that the distinguishability of quantum
states can be measured by the quantum relative entropy.) Since the
relative entropy is jointly convex in its arguments \cite{NC}, it
follows that the Holevo quantity of the periodic channel
$\Omega_{Dep}$ is also convex.

Therefore, by (\ref{perChannel}),
\be \chi\left( \left\{
p_j^{(n)}, \Omega_{per}^{(n)} \left( \rho_j^{(n)} \right)
\right\}\right) \leq \frac{1}{L} \sum_{i=0}^{L-1}  \chi \left(
\left\{ p_j^{(n)}, \Psi_i^{(n)}\left( \rho_j^{(n)} \right) \right\}
\right). \ee
Using equation (\ref{ineqEn}) we thus have
\be
\frac{1}{L} \sum_{i=0}^{L-1}  \chi^* \left( \Delta_{\lambda_i}
\right) + \frac{\epsilon}{2} \leq \frac{1}{nL} \sum_{i=0}^{L-1}
\chi \left( \left\{ p_j^{(n)}, \Psi_i^{(n)}\left( \rho_j^{(n)}
\right) \right\} \right). \ee
It follows that there is an index
$i$ such that \bea\label{ineq-i} \frac{1}{L} \sum_{i=0}^{L-1}
\chi^* \left( \Delta_{\lambda_i} \right) + \frac{\epsilon}{2} \leq
\frac{1}{n} \chi \left( \left\{ p_j^{(n)}, \Psi_i^{(n)}\left(
\rho_j^{(n)} \right) \right\} \right). \eea
But equation
(\ref{add}) implies that \be \chi \left( \left\{ p_j^{(n)},
\Psi_i^{(n)}\left( \rho_j^{(n)} \right) \right\} \right) \leq
\frac{n}{L} \sum_{i=0}^{L-1}  \chi^* \left( \Delta_{\lambda_i}
\right). \ee
Therefore the inequalities (\ref{ineq-i}) and hence
the assumption made in equation (\ref{assumption}) cannot hold,
and \bea\label{lesseq} C \left( \Omega_{per} \right) \leq C_p
\left( \Omega_{per} \right). \eea
The above equation together with equation (\ref{greatEq}) yields the required result.
\end{proof}

\section{The classical capacity of a convex combination of memoryless channels\protect\\}

In \cite{DD07CC} the product state capacity of a convex
combination of memoryless channels was determined. Given a finite
collection of memoryless channels $\Phi_1, \dots,\Phi_M$ with
common input Hilbert space $\cal H$ and output Hilbert space $\cal
K$, a convex combination of these channels is defined by the map
\be \Phi^{(n)}\left(\rho^{(n)} \right)= \sum_{i=1}^M \gamma_i
\,\Phi_i^{\otimes n}(\rho^{(n)}), \ee where $\gamma_i,\
(i=1,\dots,M)$ is a probability distribution over the channels
$\Phi_1,\dots,\Phi_M$. Thus, a given input state $\rho^{(n)} \in
{\mathcal{B}}({\cal H}^{\otimes n})$ is sent down one of the
memoryless channels with probability $\gamma_i$. This introduces
long-term memory, and as a result the (product-state) capacity of
the channel $\Phi^{(n)}$ is no longer given by the supremum of the
Holevo quantity. Instead, it was proved in \cite{DD07CC} that the
product-state capacity is given by \be C_p(\Phi) =
\sup_{\{p_j,\rho_{j}\}} \left[ \bigwedge_{i=1}^M
\chi(\{p_j,\Phi_i(\rho_j)\}) \right]. \label{convexcap} \ee

Let  $\Delta_{\lambda_i}$ be depolarizing channels with parameters
$\lambda_i$ as above, and $\Phi_{rand}$ denote the channel whose
memoryless channel branches are given by $\Lambda_i^{(n)}$ where
\be \Lambda_i^{(n)} = \Delta_{\lambda_i}^{\otimes n}. \ee  Since
the capacity of the depolarizing channel decreases with the error
parameter the product state capacity of $\Phi_{rand}$ is given by
\be C_p (\Phi_{rand}) = \bigwedge_{i=1}^M \chi^*
(\Delta_{\lambda_i}) = \chi^* \left(\bigvee_{i=1}^M \lambda_i
\right). \ee We aim to prove the following theorem.

\begth The classical capacity of a convex combination of
depolarizing channels is equal to its product state capacity \bes
C \left(\Phi_{rand} \right) = C_p \left( \Phi_{rand}\right). \ees
\enth

\bpr According to \cite{DD07CC} the classical capacity of this
channel can be written as follows \be \label{ClassCapCC}
C\left( \Phi_{rand} \right)  =  \lim_{ n \rightarrow
\infty }\; \frac{1}{n} \, \sup_{\{ p_j^{(n)}, \rho_j^{(n)} \}}
\bigwedge_{i=1}^M \chi \left( \left\{ p_j^{(n)}, \Lambda_i^{(n)}
\left(\rho_j^{(n)}\right) \right\} \right). \ee Suppose that \be
\label{assumptionConvex} C\left( \Phi_{rand} \right)
\geq  \bigwedge_{i=1}^M  \chi^*(\Delta_{\lambda_i}) + \epsilon,
\ee for some $\epsilon >0$.

Then $\exists\, n_0$, such that if $n \geq n_0$, then
\bea
\frac{1}{n} \, \sup_{\{ p_j^{(n)},  \rho_j^{(n)} \}}
\bigwedge_{i=1}^M \chi \left( \left\{ p_j^{(n)}, \Lambda_i^{(n)}
\left(\rho_j^{(n)} \right)  \right\} \right) &\geq&
\bigwedge_{i=1}^M \chi^*(\Delta_{\lambda_i}) \non \\ &+& \epsilon.
\eea
Hence, for $n \geq n_0$ there exists an ensemble $\{p_j^{(n)}, \,
\rho_j^{(n)}\}$ such that \be \frac{1}{n} \bigwedge_{i=1}^M \chi
\left( \left\{ p_j^{(n)},\Lambda_i^{(n)} \left(\rho_j^{(n)}\right)
\right\} \right) \geq \bigwedge_{i=1}^M \chi^*(\Delta_{\lambda_i})
+ \epsilon. \ee But King \cite{King03} proved that the product
state capacity of the depolarizing channel is equal to its
classical capacity, therefore \be\label{Additivity} \chi^* \left(
\Lambda_i^{(n)} \right) = n \, \chi^* \left( \Delta_{\lambda_i}
\right). \ee
In other words, $\chi \left( \left\{ p_j^{(n)},
\Lambda_i^{(n)} \left(\rho_j^{(n)}\right) \right\} \right)$ is
bounded above by $\chi^*\left(\Delta_{\lambda_i}\right)$. Now,
if $i_0$ is such that \be \bigwedge_{i=1}^M
\chi^*(\Delta_{\lambda_i}) = \chi^*(\Delta_{\lambda_{i_0}}), \ee
then
\bea \frac{1}{n} \bigwedge_{i=1}^M \chi \left( \left\{
p_j^{(n)}, \Lambda_i^{(n)}\left(\rho_j^{(n)}\right) \right\}
\right) &\leq& \chi \left( \left\{ p_j^{(n)}, \Lambda_{i_0}^{(n)}
\left(\rho_j^{(n)}\right) \right\} \right) \non \\ &\leq&
\chi^*(\Delta_{\lambda_{i_0}}). \eea
Therefore \be \frac{1}{n}
\bigwedge_{i=1}^M \chi \left( \left\{ p_j^{(n)},
\Lambda_i^{(n)}\left(\rho_j^{(n)}\right)\right\} \right) \leq
\bigwedge_{i=1}^M \chi^*(\Delta_{\lambda_i}). \ee
This contradicts the assumption made by equation (\ref{assumptionConvex}) and
therefore \be \label{lessEQ} C\left(
\Phi_{rand}\right) \leq \bigwedge_{i=1}^M
\chi^*(\Delta_{\lambda_i}) = C_p \left( \Phi_{rand}\right). \ee On
the other hand, it is clear that $ C
\left(\Phi_{rand}\right) \geq C_p\left(\Phi_{rand}\right), $ and
therefore $ C\left( \Phi_{rand}\right) = C_p \left(
\Phi_{rand}\right). $ \epr

\textbf{Remark.} \emph{Note that, in contrast to the proof of
Theorem 1, the proof above does not rely on the invariance of the
maximizing ensemble of the depolarizing channel. The proof uses
the additivity of the Holevo quantity of the depolarizing channel
(see Eqn. (\ref{Additivity})) and the result can therefore be
generalized to channels for which the additivity of the Holevo
capacity has been proved.}


\end{document}